\documentstyle[epsf]{l-aa}

\begin{document}

\thesaurus{05 (08.11.1; 08.06.2; 08.16.5; 08.12.1; 05.01.1)}

\title{New proper motions of pre-main sequence stars in
       Taurus-Auriga}

\author{S.\ Frink \inst{1} \and S.\ R\"oser \inst{1} \and
  R.\ Neuh\"auser \inst{2} \and M.F.\ Sterzik \inst{2}}
\offprints{S.\ Frink, e-mail: sabine@relay.ari.uni-heidelberg.de}
\institute{Astronomisches Rechen-Institut Heidelberg,
    M\"onchhofstra{\ss}e 12-14, D-69120 Heidelberg, Germany \and
    Max-Planck-Institut f\"ur extraterrestrische Physik,
    Giessenbachstra{\ss}e 1, D-85740 Garching, Germany}

\date{Received ; accepted }

\maketitle

\begin{abstract}
We present proper motions of 72~T~Tauri stars located in the
central region of Taurus-Auriga (Tau-Aur). These proper motions are taken
{}from a new proper motion catalogue called STARNET.
Our sample comprises 17~classical T~Tauri stars (CTTS) and 55~weak-line
T~Tauri stars (WTTS), most of the latter discovered by ROSAT.
53~stars had no proper motion measurement before.

Kinematically, 62 of these stars are members of the association.
A velocity dispersion of less than 2-3\,km\,s$^{-1}$ is found which is
dominated by the errors of the proper motions. This velocity dispersion
correlates with a spread in distances.

Furthermore we present proper motions of 58~stars located in a region
just south of the Taurus molecular clouds and compare the kinematics
of the youngest stars in this sample (younger than $3.5\cdot10^{7}$\,yrs)
with the kinematics of the pre-main sequence stars (PMS) in the 
Taurus-Auriga association.
{}From a comparison of the space velocities we find that the stars in the 
central region
of Tau-Aur are kinematically different from the stars in the southern part.

Among the stars with large proper motions far off the Taurus mean motion
we find 2~Pleiades candidates and 7~possible Pleiades runaway stars.

\end{abstract}

\section{Introduction}
The T-association Taurus-Auriga (Tau-Aur) is a large star forming region
with several hundreds of pre-main sequence stars (PMS) known.
Its distance of only about 140 pc (Elias 1978, Kenyon et al.\ 1994)
makes it an ideal object for studying the properties of the
cloud resulting from ongoing star formation, e.g. its stellar content, age,
distribution and motion of the stars and so on.

Before the launch of ROSAT about 150 T Tauri stars which are associated with
Taurus-Auriga have been known, most of them classical T Tauri stars
(CTTS). The ROSAT All-Sky Survey (RASS) and some deep ROSAT pointings
have revealed so far 4 new CTTS and 82 weak-line T Tauri stars
(WTTS) in the central region of Taurus-Auriga, between
4$^{h}$ and 5$^{h}$ in right ascension and
+15$^{\mbox{\scriptsize o}}$ and +34$^{\mbox{\scriptsize o}}$
in declination (Wichmann et al.\ 1996, Strom \& Strom 1994,
Carkner et al.\ 1996). Both CTTS and WTTS show the
characteristic strong Li\,{\sc i} $\lambda$\,6708~\AA\ absorption line,
but in contrast to CTTS, WTTS are stronger X-ray-emitters
(Neuh\"auser et al. 1995a) and so easier
to detect by ROSAT. WTTS however lack the strong H$\alpha$ emission,
Ca\,{\sc ii} emission and IR-excess of the CTTS, most of which were
discovered by objective prism surveys.

In addition to these new TTS discovered in the central region of
Taurus-Auriga pre-main sequence stars have also been found in a large
area south of the Taurus-Auriga dark clouds (Neuh\"auser et al.\ 1995b,
Magazz\`u et al.\ 1997, Neuh\"auser et al.\ 1997).

Proper motions of all these stars are required to kinematically investigate 
their membership to the association and to address the question whether 
it is possible that the stars
outside the commonly adopted boundaries of Taurus-Auriga
(based on CO surveys (Ungerechts \& Thaddeus 1987)\,)
have formed near the centre of the region or rather near their present
locations. Until now proper motions have been known only for about 70 mostly
classical T Tauri stars in selected areas inside the large region
(Jones \& Herbig 1979, Walter et al.\ 1987, Hartmann et al.\ 1991,
Gomez et al.\ 1992).

\begin{table*}
\begin{center}
\caption[]{\label{liste} STARNET proper motions for 34 stars which were 
  known to be of PMS nature before ROSAT. In the second column references
  for previous measured proper motions are given: (1) Jones \& Herbig 1979,
  (2) Walter et al.\ 1987, (3) Hartmann et al.\ 1991, (4) Gomez et al.\ 1992.
  The last column indicates
  whether it is a classical (c) or a weak-line (w) T Tauri star.
  Note that the values for SAO 76411 A are quoted from the PPM as there is
  no corresponding entry in STARNET. Remarks after the name of a star refer
  to stars not regarded as members of the Tau-Aur association and discussed 
  separately in Sect.~\protect\ref{starnet-discussion} 
  ($^{\rm (P)}$~Pleiades candidates, $^{\rm (h)}$~other high proper motion
  stars).}
\begin{tabular}{l@{\,}crr
                r@{}c@{}r@{}c@{}r@{}c@{}l
                r@{}c@{}r@{}c@{}r@{}c@{}l
                rrrrrc}
\hline\noalign{\smallskip}
& & \multicolumn{2}{c}{GSC No./} & \multicolumn{7}{c}{RA} &
\multicolumn{7}{c}{DEC} & \multicolumn{1}{c}{$\mbox{m}_{\mbox{\tiny V}}$} & 
\multicolumn{1}{c}{$\mu_{\alpha}$} 
& \multicolumn{1}{c}{$\mu_{\delta}$} & 
\multicolumn{1}{l}{\raisebox{1ex}[-1ex]{$\sigma\mu_{\alpha}$}
\raisebox{-1ex}[1ex]{\scriptsize \hspace*{-0.6cm}$\cos\delta$}} &
\multicolumn{1}{c}{$\sigma\mu_{\delta}$} & \\
\raisebox{1.5ex}[-1.5ex]{object} & \raisebox{1.5ex}[-1.5ex]{ref.} &
\multicolumn{2}{c}{PPM No.} & 
\multicolumn{7}{c}{($\alpha_{2000}$)} & 
\multicolumn{7}{c}{($\delta_{2000})$} & \multicolumn{1}{c}{[mag]} & 
\multicolumn{2}{c}{[mas/y]} & 
\multicolumn{2}{c}{[mas/y]} & \raisebox{1.5ex}[-1.5ex]{TTS} \\
\noalign{\smallskip}
\hline\noalign{\smallskip}
NTTS 032641+2420$\,^{\rm (P)}$& & 1802 & 1190 &\hphantom{1}3 &$^{h}$ & 29 &$^{m}$& 38. &$\!\!^{s}$ & 39 & +24 &$^{\mbox{\scriptsize o}}$ & 30 &\arcmin& 37. &$\!\!\arcsec$& 6 & 11.69  &   31 &   -43  &   3.4 &   3.4 & w \\
NTTS 034903+2431$\,^{\rm (P)}$& & 1804 &  123 &  3 & & 52 & &  2. & & 26 & +24 & & 39 & & 47. & & 6 & 11.17  &   27 &   -51  &   4.8 &   4.8 & w \\
SAO 76411 A$\,^{\rm (P)}$ &  & \multicolumn{2}{c}{93187} & 4 & &  2 & & 53. & & 54 & +22 & & 08 & & 11. & & 7 &  9.30  &   43 &   -58  &   3.6 &   3.7 & w \\
SAO 76428          &     & 1262 &  421 &  4 & &  4 & & 28. & & 48 & +21 & & 56 & &  4. & & 0 &  9.33  &    4 &   -27  &   4.4 &   4.4 & w \\
V773 Tau           & (1) & 1827 & 1236 &  4 & & 14 & & 12. & & 94 & +28 & & 12 & & 12. & & 3 & 10.39  &   11 &   -21  &   4.9 &   5.0 & c \\[0.8ex]
FM Tau             &     & 1827 & 1032 &  4 & & 14 & & 13. & & 59 & +28 & & 12 & & 48. & & 9 & 13.75  &    3 &   -39  &   5.9 &   5.9 & c \\
CY Tau             & (1) & 1827 &  212 &  4 & & 17 & & 33. & & 79 & +28 & & 20 & & 46. & & 8 & 12.98  &   17 &   -20  &   5.9 &   5.9 & c \\
V410 Tau           & (1) & 1827 &    8 &  4 & & 18 & & 31. & & 16 & +28 & & 27 & & 15. & & 9 & 10.37  &   17 &   -31  &   5.0 &   5.0 & w \\
BP Tau             & (1) & 1827 &  554 &  4 & & 19 & & 15. & & 85 & +29 & &  6 & & 26. & & 9 & 11.59  &   11 &   -29  &   3.9 &   3.9 & c \\
RY Tau             &     & 1828 &  129 &  4 & & 21 & & 57. & & 43 & +28 & & 26 & & 35. & & 4 & 10.09  &   16 &   -28  &   3.9 &   3.9 & c \\[0.8ex]
HDE 283572         & (2) & 1828 &  481 &  4 & & 21 & & 58. & & 86 & +28 & & 18 & &  6. & & 3 &  8.70  &   11 &   -31  &   3.6 &   3.6 & w \\
T Tau N            & (1) & 1272 &  470 &  4 & & 21 & & 59. & & 40 & +19 & & 32 & &  6. & & 5 &  8.95  &   10 &   -13  &   4.4 &   4.4 & c \\
DF Tau             & (3) & 1820 &  525 &  4 & & 27 & &  2. & & 79 & +25 & & 42 & & 23. & & 1 & 11.82  &    0 &    -6  &   3.9 &   3.9 & c \\
DG Tau             & (3) & 1820 &  330 &  4 & & 27 & &  4. & & 70 & +26 & &  6 & & 16. & & 6 & 11.59  &    7 &   -10  &   3.7 &   3.7 & c \\
NTTS 042417+1744   & (4) & 1269 &  913 &  4 & & 27 & & 10. & & 57 & +17 & & 50 & & 42. & & 3 & 10.03  &    3 &   -17  &   3.3 &   3.3 & w \\[0.8ex]
UX Tau A           & (4) & 1269 &  225 &  4 & & 30 & &  3. & & 90 & +18 & & 13 & & 49. & & 6 & 10.17  &   -2 &   -11  &   3.7 &   3.7 & w \\
DK Tau             & (3) & 1833 &   37 &  4 & & 30 & & 44. & & 27 & +26 & &  1 & & 25. & & 0 & 12.01  &   12 &   -14  &   5.0 &   5.0 & c \\
L1551-51           & (4) & 1270 & 1195 &  4 & & 32 & &  9. & & 30 & +17 & & 57 & & 22. & & 6 & 11.69  &   12 &   -20  &   4.3 &   4.3 & w \\
V827 Tau           &     & 1270 & 1108 &  4 & & 32 & & 14. & & 56 & +18 & & 20 & & 14. & & 8 & 12.00  &   10 &   -16  &   4.0 &   4.0 & w \\
V826 Tau           & (4) & 1270 &  604 &  4 & & 32 & & 15. & & 84 & +18 & &  1 & & 38. & & 7 & 11.81  &   11 &   -23  &   4.3 &   4.3 & w \\[0.8ex]
GG Tau             &     & 1270 &  897 &  4 & & 32 & & 30. & & 32 & +17 & & 31 & & 40. & & 3 & 11.85  &   11 &   -28  &   3.9 &   3.9 & c \\
V807 Tau           &     & 1829 &  214 &  4 & & 33 & &  6. & & 62 & +24 & &  9 & & 54. & & 9 & 11.24  &   10 &   -21  &   3.7 &   3.7 & c \\
V830 Tau           & (3) & 1833 &  843 &  4 & & 33 & &  9. & & 99 & +24 & & 33 & & 42. & & 8 & 12.03  &  -14 &   -32  &   4.7 &   4.7 & w \\
NTTS 043124+1824   & (4) & 1270 &  232 &  4 & & 34 & & 18. & & 01 & +18 & & 30 & &  6. & & 6 & 12.70  &   -7 &    -8  &   3.9 &   3.9 & w \\
NTTS 043220+1815   &     & 1270 & 1331 &  4 & & 35 & & 14. & & 20 & +18 & & 21 & & 35. & & 5 & 10.85  &   -3 &   -15  &   3.3 &   3.3 & w \\[0.8ex]
DN Tau             & (1) & 1829 &   26 &  4 & & 35 & & 27. & & 37 & +24 & & 14 & & 58. & & 8 & 12.17  &    3 &   -21  &   3.6 &   3.6 & c \\
LkCa 14$\,^{\rm (h)}$   &     & 1834 &  177 &  4 & & 36 & & 18. & & 91 & +25 & & 43 & &  1. & & 4 & 11.61  & -161 &    97  &   4.6 &   4.6 & w \\
LkCa 15            &     & 1278 &  193 &  4 & & 39 & & 17. & & 76 & +22 & & 21 & &  3. & & 7 & 11.65  &    0 &    -7  &   4.7 &   4.7 & c \\
DS Tau             &     & 1843 &  937 &  4 & & 47 & & 48. & & 42 & +29 & & 25 & & 12. & & 2 & 11.21  &  -18 &   -19  &   4.9 &   4.9 & c \\
UY Aur             & (1) & 2387 &  982 &  4 & & 51 & & 47. & & 38 & +30 & & 47 & & 13. & & 2 & 11.75  &    6 &   -26  &   5.1 &   5.1 & c \\[0.8ex]
LkCa 19            &     & 2387 &  637 &  4 & & 55 & & 36. & & 97 & +30 & & 17 & & 55. & & 2 & 10.51  &    4 &   -18  &   4.2 &   4.2 & w \\
SU Aur             & (1) & 2387 &  977 &  4 & & 55 & & 59. & & 24 & +30 & & 34 & &  1. & & 5 &  8.37  &  -18 &   -28  &   4.2 &   4.2 & c \\
NTTS 045251+3016   &     & 2387 &  535 &  4 & & 56 & &  2. & & 04 & +30 & & 21 & &  3. & & 6 & 11.37  &    5 &   -25  &   4.2 &   4.2 & w \\
RW Aur A           & (1) & 2389 &  955 &  5 & &  7 & & 49. & & 50 & +30 & & 24 & &  5. & & 0 & 10.47  &  -11 &   -27  &   4.3 &   4.3 & c \\
\noalign{\smallskip}\hline\\[0.8ex]
\end{tabular}
\end{center}
\end{table*}

\begin{table*}
\begin{center}
\caption[]{\label{wich} All new ROSAT-discovered PMS stars (all WTTS) in the 
  central region of Tau-Aur which we could identify in STARNET. The entries 
  quoted for HD 283798 are from the PPM as the proper motions from PPM and 
  STARNET deviate significantly for this star, probably due to a 
  misidentification. Because BD\,+17 724 B is not present in STARNET we added
  its proper motion from the PPM. Remarks after the name of a star refer to
  stars with proper motions investigated in Table~\protect\ref{test}.}
\begin{tabular}{lrr
                r@{}c@{}r@{}c@{}r@{}c@{}l
                r@{}c@{}r@{}c@{}r@{}c@{}l
                rrrrr}
\hline\noalign{\smallskip}
& \multicolumn{2}{c}{GSC No./} & \multicolumn{7}{c}{RA} &
\multicolumn{7}{c}{DEC} & \multicolumn{1}{c}{$\mbox{m}_{\mbox{\tiny V}}$} & 
\multicolumn{1}{c}{$\mu_{\alpha}$} 
& \multicolumn{1}{c}{$\mu_{\delta}$} & 
\multicolumn{1}{l}{\raisebox{1ex}[-1ex]{$\sigma\mu_{\alpha}$}
\raisebox{-1ex}[1ex]{\scriptsize \hspace*{-0.6cm}$\cos\delta$}} &
\multicolumn{1}{c}{$\sigma\mu_{\delta}$} \\
\raisebox{1.5ex}[-1.5ex]{object} & 
\multicolumn{2}{c}{PPM No.} & 
\multicolumn{7}{c}{($\alpha_{2000}$)} & 
\multicolumn{7}{c}{($\delta_{2000})$} & \multicolumn{1}{c}{[mag]} & 
\multicolumn{2}{c}{[mas/y]} & 
\multicolumn{2}{c}{[mas/y]} \\
\noalign{\smallskip}
\hline\noalign{\smallskip}
HD 285281        &  1258  &   894 & \hphantom{1}4 & $^{h}$ &  0 & $^{m}$ & 31. &$\!\!^{s}$ & 07 & +19 & $^{\mbox{\scriptsize o}}$ & 35 & $^{'}$ & 20. & $\!\!^{''}$ & 6 &  9.98 &  4 & -16 & 3.9 & 3.9 \\
RXJ 0403.4+1725$\,^{\rm (h)}$  &  1254  &   309 &  4 & &  3 & & 24. & & 85 & +17 & & 24 & & 26. & & 1 & 11.47  &  -66 &   -15  &   3.6 &   3.3 \\
RXJ 0405.2+2632  &  1822  &  1383 &  4 & &  5 & & 12. & & 33 & +26 & & 32 & & 43. & & 8 & 11.22  &   18 &   -22  &   4.0 &   4.0 \\
RXJ 0405.3+2009  &  1258  &   338 &  4 & &  5 & & 19. & & 60 & +20 & &	9 & & 25. & & 0 & 10.20  &    7 &   -22  &   3.6 &   3.6 \\
HD 284135        &  1814  &   409 &  4 & &  5 & & 40. & & 57 & +22 & & 48 & & 11. & & 7 &  9.26  &    2 &   -24  &   3.0 &   3.0 \\[0.8ex]
HD 284149        &  1258  &   257 &  4 & &  6 & & 38. & & 81 & +20 & & 18 & & 10. & & 6 &  9.47  &    7 &   -22  &   2.9 &   2.9 \\
RXJ 0406.8+2541  &  1818  &   144 &  4 & &  6 & & 51. & & 34 & +25 & & 41 & & 28. & & 4 & 11.30  &   11 &   -21  &   4.2 &   4.2 \\
RXJ 0407.9+1750$\,^{\rm (P)}$  &  1254  &   785 &  4 & &  7 & & 53. & & 99 & +17 & & 50 & & 26. & & 0 & 11.16  &   18 &   -49  &   3.9 &   3.9 \\
RXJ 0409.2+2901$\,^{\rm (P)}$  &  1826  &   877 &  4 & &  9 & &  9. & & 81 & +29 & &	1 & & 29. & & 8 & 10.11  &   39 &   -37  &   3.5 &   3.5 \\
RXJ 0412.8+2442  &  1819  &   498 &  4 & & 12 & & 51. & & 22 & +24 & & 41 & & 43. & & 9 & 11.66  &   11 &   -19  &   5.7 &   5.7 \\[0.8ex]
HD 285579$\,^{\rm (P)}$        &  1251  &   201 &  4 & & 12 & & 59. & & 87 & +16 & & 11 & & 47. & & 6 & 10.74  &   13 &   -50  &   3.8 &   3.8 \\
RXJ 0415.4+2044  &  1263  &  1027 &  4 & & 15 & & 22. & & 95 & +20 & & 44 & & 16. & & 9 & 10.35  &    8 &   -17  &   4.6 &   4.6 \\
RXJ 0420.4+3123  &  2371  &   740 &  4 & & 20 & & 24. & & 13 & +31 & & 23 & & 23. & & 8 & 12.18  &    1 &   -15  &   7.9 &   7.9 \\
BD\,+26 718      &  1824  &   592 &  4 & & 24 & & 48. & & 18 & +26 & & 43 & & 16. & & 1 & 11.46  &   15 &   -20  &   5.0 &   5.0 \\
BD\,+26 718 B    &  1824  &   183 &  4 & & 24 & & 49. & & 09 & +26 & & 43 & &  9. & & 5 & 10.47  &   13 &   -27  &   5.0 &   5.0 \\[0.8ex]
BD\,+17 724 B & \multicolumn{2}{c}{119907} & 4 & & 27 & &  5. & & 97 & +18 & & 12 & & 37. & & 6 &  9.50 &    3 &  -7 & 3.8 & 3.8 \\
RXJ 0430.8+2113  &  1277  &   574 &  4 & & 30 & & 49. & & 15 & +21 & & 14 & & 10. & & 5 &  9.85  &   22 &   -26  &   4.4 &   4.4 \\
HD 284496        &  1277  &  1238 &  4 & & 31 & & 16. & & 85 & +21 & & 50 & & 25. & & 4 & 10.40  &   -5 &   -12  &   4.4 &   4.4 \\
RXJ 0432.7+1853  &  1274  &  1501 &  4 & & 32 & & 42. & & 43 & +18 & & 55 & & 10. & & 0 & 10.54  &   -3 &   -18  &   3.5 &   3.5 \\
RXJ 0433.7+1823  &  1270  &   230 &  4 & & 33 & & 42. & & 00 & +18 & & 24 & & 27. & & 4 & 11.93  &  -14 &    -9  &   3.3 &   3.3 \\[0.8ex]
RXJ 0437.5+1851$\,^{\rm (P)}$  &  1274  &  1515 &  4 & & 37 & & 26. & & 87 & +18 & & 51 & & 25. & & 2 & 10.73  &   16 &   -52  &   3.4 &   3.4 \\
HD 285957        &  1266  &  1195 &  4 & & 38 & & 39. & & 01 & +15 & & 46 & & 12. & & 9 &  9.86  &   -5 &   -33  &   5.2 &   5.2 \\
RXJ 0439.4+3332A$\,^{\rm (P)}$ &  2378  &  1232 &  4 & & 39 & & 25. & & 48 & +33 & & 32 & & 44. & & 8 & 11.17  &   30 &   -40  &   9.2 &   9.2 \\
HD 283798 & \multicolumn{2}{c}{93702} & 4 & & 41 & & 55. & & 16 & +26 & & 58 & & 49. & & 3 &  9.80 &   -3 & -26  & 2.2 & 2.1 \\
RXJ 0444.9+2717  &  1839  &   643 &  4 & & 44 & & 54. & & 40 & +27 & & 17 & & 45. & & 6 &  9.00  &   -2 &   -18  &   5.1 &   5.1 \\[0.8ex]
HD 30171         &  1267  &   425 &  4 & & 45 & & 51. & & 29 & +15 & & 55 & & 49. & & 2 &  8.95  &   10 &   -28  &   4.2 &   4.2 \\
RXJ 0448.0+2755  &  1839  &  1278 &  4 & & 48 & &  0. & & 41 & +27 & & 56 & & 19. & & 8 & 12.35  &  -10 &    -2  &   6.3 &   6.3 \\
RXJ 0450.0+2230  &  1292  &   639 &  4 & & 50 & &  0. & & 18 & +22 & & 29 & & 57. & & 7 & 11.07  &   -5 &   -14  &   3.3 &   3.3 \\
RXJ 0452.8+1621  &  1280  &   559 &  4 & & 52 & & 50. & & 14 & +16 & & 22 & &  9. & & 1 & 11.31  &    1 &   -23  &   4.8 &   4.8 \\
RXJ 0453.0+1920  &  1288  &   790 &  4 & & 52 & & 57. & & 06 & +19 & & 19 & & 50. & & 1 & 12.07  &   -3 &   -23  &   3.9 &   3.9 \\[0.8ex]
HD 31281         &  1284  &  1193 &  4 & & 55 & &  9. & & 60 & +18 & & 26 & & 30. & & 6 &  9.12  &   -9 &   -28  &   3.8 &   3.8 \\
RXJ 0455.8+1742  &  1284  &   522 &  4 & & 55 & & 47. & & 64 & +17 & & 42 & &  1. & & 6 & 11.16  &  -12 &   -28  &   3.9 &   3.9 \\
HD 286179        &  1281  &  1215 &  4 & & 57 & &  0. & & 62 & +15 & & 17 & & 52. & & 6 & 10.06  &   -8 &   -24  &   5.1 &   5.1 \\
RXJ 0457.1+3142  &  2388  &   857 &  4 & & 57 & &  6. & & 52 & +31 & & 42 & & 50. & & 0 & 10.06  &   -2 &   -20  &   4.4 &   4.4 \\
RXJ 0457.2+1524  &  1281  &  1288 &  4 & & 57 & & 17. & & 64 & +15 & & 25 & &  9. & & 1 & 10.11  &   -1 &   -23  &   4.9 &   4.9 \\[0.8ex]
RXJ 0457.5+2014  &  1289  &   513 &  4 & & 57 & & 30. & & 63 & +20 & & 14 & & 28. & & 5 & 10.89  &   -9 &   -34  &   3.6 &   3.6 \\
RXJ 0458.7+2046  &  1293  &  2396 &  4 & & 58 & & 39. & & 71 & +20 & & 46 & & 43. & & 1 & 11.65  &   -3 &   -38  &   5.7 &   5.7 \\
RXJ 0459.8+1430  &   697  &   960 &  4 & & 59 & & 46. & & 14 & +14 & & 30 & & 55. & & 1 & 11.81  &    3 &   -21  &   5.0 &   5.0 \\
\noalign{\smallskip}\hline
\end{tabular}
\end{center}
\end{table*}

\begin{table*}
\begin{center}
\caption[]{\label{neu} STARNET proper motions for stars not older than 
  $3.5\cdot10^{7}$\,yrs, $10^{8}$\,yrs old stars and stars older than
  $10^{8}$\,yrs in a
  region located somewhat south of the Taurus molecular clouds. The entries for
  the stars BD\,+07 582 and BD\,+07 582 B are from PPM as there is only one 
  entry in STARNET for both components with a quite different proper motion.
  Remarks after a star's name refer again to stars discussed in 
  Sect.~\protect\ref{starnet-discussion}.}
\begin{tabular}{lrr
                r@{}c@{}r@{}c@{}r@{}c@{}l
                r@{}c@{}r@{}c@{}r@{}c@{}l
                rrrrr}
\hline\noalign{\smallskip}
& \multicolumn{2}{c}{GSC No./} & \multicolumn{7}{c}{RA} &
\multicolumn{7}{c}{DEC} & \multicolumn{1}{c}{$\mbox{m}_{\mbox{\tiny V}}$} & 
\multicolumn{1}{c}{$\mu_{\alpha}$} 
& \multicolumn{1}{c}{$\mu_{\delta}$} & 
\multicolumn{1}{l}{\raisebox{1ex}[-1ex]{$\sigma\mu_{\alpha}$}
\raisebox{-1ex}[1ex]{\scriptsize \hspace*{-0.6cm}$\cos\delta$}} &
\multicolumn{1}{c}{$\sigma\mu_{\delta}$} \\
\raisebox{1.5ex}[-1.5ex]{object} & 
\multicolumn{2}{c}{PPM No.} & 
\multicolumn{7}{c}{($\alpha_{2000}$)} & 
\multicolumn{7}{c}{($\delta_{2000})$} & \multicolumn{1}{c}{[mag]} & 
\multicolumn{2}{c}{[mas/y]} & 
\multicolumn{2}{c}{[mas/y]} \\
\noalign{\smallskip}\hline\hline\noalign{\smallskip}
\multicolumn{22}{l}{stars not older than $3.5\cdot10^{7}$\,yrs} \\
\noalign{\smallskip}\hline\noalign{\smallskip}
RXJ 0324.4+0231    &    60  &   489 &  3 &$^{h}$ & 24 &$^{m}$& 25. &$\!\!^{s}$ & 25 & + 2 &$^{\mbox{\scriptsize o}}$ &31 &\arcmin&  1. &$\!\!\arcsec$& 1 & 12.83  &   24 &    -3  &   4.3 &   4.3 \\
RXJ 0338.3+1020    &   660  &   709 &  3 & & 38 & & 18. & & 21 & +10 & & 20 & & 16. & & 7 & 10.97  &   19 &   -25  &   3.5 &   3.5  \\
RXJ 0344.8+0359    &    68  &  1471 &  3 & & 44 & & 53. & & 16 & + 3 & & 59 & & 30. & & 4 & 12.26  &   17 &   -18  &   4.9 &   4.9  \\
RXJ 0347.9+0616    &    71  &   542 &  3 & & 47 & & 56. & & 82 & + 6 & & 16 & &  6. & & 9 & 11.06  &   15 &    -5  &   3.5 &   3.5  \\
RXJ 0348.5+0832    &   658  &   922 &  3 & & 48 & & 31. & & 47 & + 8 & & 31 & & 37. & & 7 & 10.90  &   33 &    -7  &   4.0 &   4.0  \\[0.8ex]
RXJ 0354.1+0528    &    72  &   606 &  3 & & 54 & &  6. & & 57 & + 5 & & 27 & & 23. & & 4 & 11.63  &   -9 &    -9  &   3.9 &   3.9  \\
RXJ 0354.3+0535    &    72  &   921 &  3 & & 54 & & 21. & & 27 & + 5 & & 35 & & 40. & & 8 & 10.17  &   -9 &    -5  &   3.9 &   3.9  \\
RXJ 0357.3+1258    &   665  &   150 &  3 & & 57 & & 21. & & 37 & +12 & & 58 & & 16. & & 9 & 10.89  &   21 &   -18  &   5.2 &   5.2  \\
BD\,+07 582 B$\;^{(\ast)}$ & \multicolumn{2}{c}{147111} & 4 & &  0 & &  9. & & 39 & +8 & & 18 & & 18. & & 9 & 10.80 & -22 & -15 &   9.3 &   9.3  \\
RXJ 0407.2+0113 N  &    73  &   762 &  4 & &  7 & & 16. & & 44 & + 1 & & 13 & & 14. & & 4 & 11.14  &   18 &    -4  &   3.4 &   3.4  \\[0.8ex]
RXJ 0422.9+0141    &    75  &    41 &  4 & & 22 & & 54. & & 61 & + 1 & & 41 & & 31. & & 8 & 12.29  &  -11 &    -8  &   4.1 &   4.1  \\
RXJ 0427.4+1039    &   672  &  1265 &  4 & & 27 & & 30. & & 28 & +10 & & 38 & & 48. & & 6 & 11.33  &   -2 &    -9  &   4.1 &   4.1  \\
RXJ 0427.5+0616    &    81  &  1414 &  4 & & 27 & & 32. & & 07 & + 6 & & 15 & & 51. & & 9 & 10.29  &    5 &     1  &   3.5 &   3.5  \\
RXJ 0434.3+0226    &    86  &   318 &  4 & & 34 & & 19. & & 50 & + 2 & & 26 & & 25. & & 7 & 13.01  &    3 &   -25  &   4.6 &   4.6  \\
BD\,+08 742        &   682  &   674 &  4 & & 42 & & 32. & & 14 & + 9 & &  6 & &  1. & & 4 & 10.86  &   36 &   -20  &   6.6 &   6.6  \\[0.8ex]
RXJ 0450.0+0151    &    84  &   743 &  4 & & 50 & &  4. & & 69 & + 1 & & 50 & & 42. & & 7 & 12.15  &   11 &   -12  &   4.1 &   4.1  \\
RXJ 0511.9+1112    &   702  &  1689 &  5 & & 12 & &  0. & & 29 & +11 & & 12 & & 19. & & 5 & 11.38  &    2 &    -5  &   3.6 &   3.6  \\
RXJ 0512.0+1020    &   702  &  2533 &  5 & & 12 & &  3. & & 19 & +10 & & 20 & &  6. & & 7 & 11.32  &    1 &     6  &   3.6 &   3.6  \\
\noalign{\smallskip}\hline\noalign{\smallskip}
\multicolumn{22}{l}{$\approx 10^{8}$\,yrs old stars} \\
\noalign{\smallskip}\hline\noalign{\smallskip}
RXJ 0219.7-1026    &  5282  &  2210 &  2 &$^{h}$ & 19 &$^{m}$& 47. &$\!\!^{s}$ & 39 & -10 &$^{\mbox{\scriptsize o}}$ &25 &\arcmin& 40. &$\!\!\arcsec$& 7 & 11.60  &   15 &    -2  &   3.7 &   3.7 \\
HD 15526$\,^{\rm (h)}$  &  5284  &   686 &  2 & & 29 & & 35. & & 07 & -12 & & 24 & &  9. & & 0 & 10.32  &   48 &   -13  &   3.8 &   3.8  \\
RXJ 0329.1+0118    &    57  &   485 &  3 & & 29 & &  8. & & 06 & + 1 & & 18 & &  5. & & 5 & 10.90  &    4 &    -6  &   2.9 &   2.9  \\
RXJ 0339.6+0624    &    70  &  1148 &  3 & & 39 & & 40. & & 56 & + 6 & & 24 & & 43. & & 5 & 11.70  &   -3 &    -6  &   5.5 &   5.5  \\
RXJ 0343.6+1039    &   660  &   825 &  3 & & 43 & & 40. & & 49 & +10 & & 39 & & 13. & & 7 & 10.21  &   -8 &   -32  &   3.8 &   3.8  \\[0.8ex]
BD\,+11 533$\;^{(\ast)}$&   661  &   452 &  3 & & 52 & & 24. & & 76 & +12 & & 22 & & 43. & & 2 &  9.65  &    5 &   -25  &   4.4 &   4.4  \\
BD\,+07 582$\;^{(\ast)}$& \multicolumn{2}{c}{147110} & 4 & &  0 & &  9. & & 32 & +8 & & 18 & & 13. & & 7 & 10.70 & -10 & -36 &   4.6 &   4.6  \\
RXJ 0404.4+0519    &    79  &   810 &  4 & &  4 & & 28. & & 48 & + 5 & & 18 & & 43. & & 0 & 11.18  &   10 &   -11  &   3.0 &   3.0  \\
HD 286556          &   674  &   504 &  4 & &  9 & & 51. & & 55 & +12 & &  9 & &  1. & & 9 & 12.02  &    4 &   -28  &   4.2 &   4.2  \\
RXJ 0423.5+0955    &   672  &  1156 &  4 & & 23 & & 30. & & 22 & + 9 & & 54 & & 29. & & 3 & 11.64  &  -15 &   -14  &   5.3 &   5.3  \\[0.8ex]
HD 286753          &   676  &  1123 &  4 & & 25 & & 35. & & 33 & +12 & &  9 & & 59. & & 4 & 10.40  &   30 &   -23  &   3.4 &   3.4  \\
RXJ 0442.9+0400    &    91  &   702 &  4 & & 42 & & 54. & & 72 & + 4 & &  0 & & 11. & & 6 & 11.13  &   13 &   -17  &   3.6 &   3.6  \\
\noalign{\smallskip}\hline\noalign{\smallskip}
\multicolumn{22}{l}{stars older than $10^{8}$\,yrs} \\
\noalign{\smallskip}\hline\noalign{\smallskip}
RXJ 0210.4-1308 SW$\,^{\rm (h)}$ &  5283  &  1690 &\hphantom{1}2 &$^{h}$ & 10 &$^{m}$& 25. &$\!\!^{s}$ & 82 & -13 &$^{\mbox{\scriptsize o}}$ & 7 &\arcmin& 56. &$\!\!\arcsec$& 6 & 10.48  &   55 &   -24  &   4.0 &   4.0 \\
RXJ 0212.3-1330$\,^{\rm (h)}$   &   5283 &   876 &  2 & & 12 & & 18. & & 73 & -13 & & 30 & & 42. & & 4 & 11.41  &  163 &   -81  &   4.1 &   4.1  \\
RXJ 0218.6-1004    &   5282 &    68 &  2 & & 18 & & 39. & & 56 & -10 & &  4 & &  6. & & 0 & 11.75  &   18 &   -16  &   4.4 &   4.4  \\
RXJ 0239.1-1028$\,^{\rm (h)}$    &  5288  &  1027 &  2 & & 39 & &  8. & & 77 & -10 & & 27 & & 46. & & 4 & 13.21  &   -8 &   -95  &   5.0 &   5.0  \\
RXJ 0248.3-1117    &   5289 &  1010 &  2 & & 48 & & 22. & & 21 & -11 & & 17 & & 12. & & 4 & 12.02  &   14 &   -10  &   3.8 &   3.8  \\[0.8ex]
RXJ 0309.1+0324    &     58 &   166 &  3 & &  9 & &  9. & & 89 & + 3 & & 23 & & 44. & & 0 & 10.39  &   31 &    -9  &   3.5 &   3.4  \\
RXJ 0317.9+0231$\,^{\rm (h)}$    &     59 &    24 &  3 & & 17 & & 59. & & 16 & + 2 & & 30 & & 12. & & 0 & 10.73  &  -15 &   -61  &   4.2 &   4.2  \\
RXJ 0330.7+0306 N$\,^{\rm (h)}$  &    67  &   206 &  3 & & 30 & & 43. & & 48 & + 3 & &  5 & & 46. & & 9 & 11.18  &   33 &   -85  &   3.9 &   3.9  \\
RXJ 0336.0+0846$\,^{\rm (h)}$    &   657  &   726 &  3 & & 36 & &  0. & & 28 & + 8 & & 45 & & 36. & & 7 & 12.35  &    1 &    26  &   7.4 &   7.4  \\
BD\,+12 511        &\multicolumn{2}{c}{119389}&  3 & & 49 & & 27. & & 76 & +12 & & 54 & & 43. & & 8 &  9.60  &  -42 &    45  &   4.0 &   3.9\\
\noalign{\smallskip}\hline\\[0.8ex]
\end{tabular}
\begin{tabular}{c@{$\;$}p{14cm}}
{\footnotesize $^{(\ast)}$} & 
{\footnotesize classification is doubtful (Neuh\"auser et al.\ 1997), these
stars may either be ZAMS or PMS stars}
\end{tabular}
\addtocounter{table}{-1}
\end{center}
\end{table*}
\begin{table*}
\begin{center}
\caption[]{(continued)}
\begin{tabular}{lrr
                r@{}c@{}r@{}c@{}r@{}c@{}l
                r@{}c@{}r@{}c@{}r@{}c@{}l
                rrrrr}
\hline\noalign{\smallskip}
& \multicolumn{2}{c}{GSC No./} & \multicolumn{7}{c}{RA} &
\multicolumn{7}{c}{DEC} & \multicolumn{1}{c}{$\mbox{m}_{\mbox{\tiny V}}$} & 
\multicolumn{1}{c}{$\mu_{\alpha}$} 
& \multicolumn{1}{c}{$\mu_{\delta}$} & 
\multicolumn{1}{l}{\raisebox{1ex}[-1ex]{$\sigma\mu_{\alpha}$}
\raisebox{-1ex}[1ex]{\scriptsize \hspace*{-0.6cm}$\cos\delta$}} &
\multicolumn{1}{c}{$\sigma\mu_{\delta}$} \\
\raisebox{1.5ex}[-1.5ex]{object} & 
\multicolumn{2}{c}{PPM No.} & 
\multicolumn{7}{c}{($\alpha_{2000}$)} & 
\multicolumn{7}{c}{($\delta_{2000})$} & \multicolumn{1}{c}{[mag]} & 
\multicolumn{2}{c}{[mas/y]} & 
\multicolumn{2}{c}{[mas/y]} \\
\noalign{\smallskip}\hline\hline\noalign{\smallskip}
\multicolumn{22}{l}{stars older than $10^{8}$\,yrs} \\
\noalign{\smallskip}\hline\noalign{\smallskip}
RXJ 0402.5+0552    &\hphantom{22} 79  & 729 &  4 &$^{h}$ &  2 &$^{m}$& 35. &$\!\!^{s}$ & 71 & + 5 &$^{\mbox{\scriptsize o}}$ &51 &\arcmin& 36. &$\!\!\arcsec$& 5 & 10.80  &   33 &    41  &   3.3 &   3.3  \\
RXJ 0405.5+0324    &    76  &   713 &  4 & &  5 & & 30. & & 24 & + 3 & & 23 & & 50. & & 1 & 11.49  &\hphantom{22}8 &    -5  &   4.4 &   4.4  \\
RXJ 0418.6+0143    &    74  &  1029 &  4 & & 18 & & 39. & & 24 & + 1 & & 42 & &  9. & & 6 & 12.34  &  -14 &   -39  &   4.2 &   4.2  \\
RXJ 0426.4+0957W   &    672 &  1133 &  4 & & 26 & & 26. & & 79 & + 9 & & 56 & & 59. & & 8 & 11.47  &   15 &    13  &   5.3 &   5.3  \\
BD\,+00 760$\,^{\rm (h)}$ & 75 &  1529 &  4 & & 27 & & 53. & & 09 & + 0 & & 49 & & 25. & & 6 & 9.60   &   57 &    -8  &   3.3 &   3.3  \\[0.8ex]
RXJ 0429.9+0155    &     75 &     1 &  4 & & 29 & & 56. & & 87 & + 1 & & 54 & & 47. & & 3 & 10.48  &   -3 &    -5  &   3.7 &   3.7  \\
RXJ 0435.5+0455    &     90 &   936 &  4 & & 35 & & 31. & & 55 & + 4 & & 55 & & 32. & & 3 & 9.91   &  -12 &     5  &   3.0 &   3.0  \\
BD\,+05 706        &     91 &   830 &  4 & & 41 & & 57. & & 70 & + 5 & & 36 & & 34. & & 5 & 9.62   &    5 &    -1  &   3.9 &   3.9  \\
RXJ 0442.3+0118    &    83  &   788 &  4 & & 42 & & 18. & & 59 & + 1 & & 17 & & 39. & & 6 & 11.79  &   -3 &   -29  &   3.3 &   3.3  \\
RXJ 0442.6+1018    &    686 &  1246 &  4 & & 42 & & 40. & & 87 & +10 & & 17 & & 44. & & 3 & 8.25   &   33 &   -33  &   5.2 &   5.2  \\[0.8ex]
HD 287017$\,^{\rm (P)}$ & 687 & 419 &  4 & & 44 & & 20. & & 44 & + 9 & & 41 & &  3. & & 3 & 8.86   &   52 &   -73  &   5.1 &   5.1  \\
RXJ 0445.3+0914    &    683 &   282 &  4 & & 45 & & 23. & & 81 & + 9 & & 13 & & 48. & & 6 & 11.75  &   -1 &    -7  &   4.6 &   4.6  \\
RXJ 0448.0+0738$\,^{\rm (P)}$ & 683 & 661 &  4 & & 48 & &  0. & & 89 & + 7 & & 37 & & 56. & & 2 & 11.15  &   25 &   -59  &   4.2 &   4.2  \\
RXJ 0515.3+1221    &   707  &  1311 &  5 & & 15 & & 20. & & 57 & +12 & & 21 & & 13. & & 5 & 11.59  &   24 &   -31  &   4.5 &   4.5  \\
RXJ 0523.5+1005    &    704 &  2521 &  5 & & 23 & & 33. & & 72 & +10 & &  4 & & 29. & & 2 & 11.23  &   -4 &   -33  &   4.4 &   4.4  \\[0.8ex]
RXJ 0523.9+1101$\,^{\rm (h)}$ & 704 &  2073 &  5 & & 23 & & 57. & & 05 & +11 & &  0 & & 55. & & 7 & 10.83  &    2 &  -115  &   3.4 &   3.4  \\
RXJ 0528.4+1213    &    708 &  2137 &  5 & & 28 & & 25. & & 70 & +12 & & 12 & & 36. & & 1 & 11.50  &   -6 &    -2  &   4.4 &   4.4  \\
RXJ 0530.9+1227    &    709 &  1637 &  5 & & 30 & & 57. & & 20 & +12 & & 27 & & 26. & & 4 & 10.69  &    6 &   -24  &   5.4 &   5.4  \\
\noalign{\smallskip}\hline\\[0.8ex]
\end{tabular}
\end{center}
\end{table*}

\begin{figure}[t]
\epsfxsize=8.5cm
\leavevmode
\epsffile{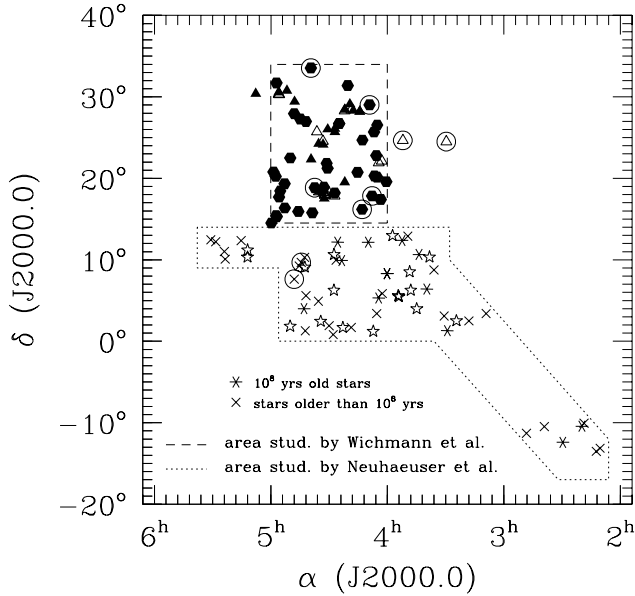}
\caption[]{\label{pos} Positions of the stars in
  Tables~\ref{liste}-\ref{neu}. Stars of Table~\ref{liste}
  (TTS known prior to ROSAT) are marked with triangles
  (CTTS filled, WTTS open),
  stars of Table~\ref{wich} (WTTS studied by Wichmann et al.\ 1996)
  with filled hexagons and the youngest stars of Table~\ref{neu} 
  (stars studied by
  Neuh\"auser et al.\ 1995b, 1997 and Magazz\`u et al.\ 1997) with open stars.
  $10^{8}$\,yrs old stars and stars older than $10^{8}$\,yrs of 
  Table~\ref{neu} are marked with centered stars and
  crosses, respectively. A circle around a symbol indicates possible
  Pleiades members, see Sect.~\ref{starnet-discussion}.
  The regions studied by those authors are indicated, too.}
\end{figure}

\begin{figure}[t]
\epsfxsize=8.5cm
\leavevmode
\epsffile{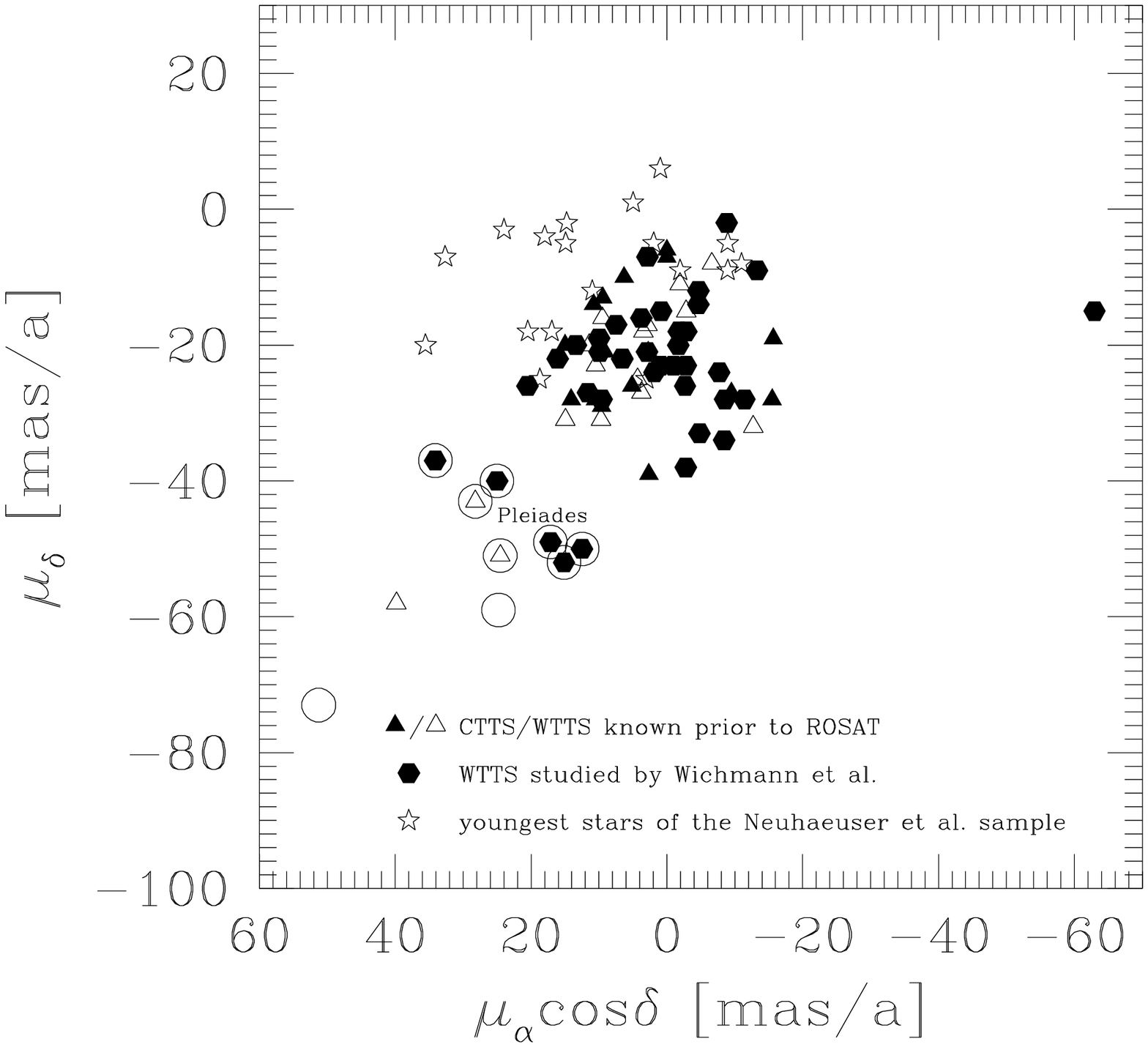}
\caption[]{\label{pm} Proper motions of all the stars in the centre 
  (Tables~\ref{liste} and \ref{wich}) and of the youngest stars in the south
  (upper part of Table~\ref{neu}). The coding of the different stars is the
  same as in Fig.~\ref{pos}. Note that the proper motion of LkCa~14 is too
  large to be shown in this diagram (it maybe wrong anyway, see
  Sect.~\ref{starnet-discussion}).}
\end{figure}

\section{Data}
To cover the whole region a deep all-sky catalogue of proper motions is
required.
The proper motions we present in the following sections are taken from
STARNET (R\"oser 1996), which is a large proper motion catalogue
well suited for our purposes. It contains about 4.3 million stars
(most stars down to V=11.5\,mag and some even fainter) with
proper motions derived from plates with nearly 100 years epoch difference
(Astrographic Catalogue (AC) and HST Guide Star Catalogue (GSC 1.2,
R\"oser et al.\ 1996)\,).
The average accuracy of the proper motions is about 5\,mas/y, which
corresponds to an error of about 3\,km\,s$^{-1}$ at the distance of Tau-Aur.

For brighter stars the PPM catalogue is a suitable source of proper
motions. It contains some 400 000 stars down to about V=10.0 mag.
On the northern hemisphere the measurements used in PPM have a smaller
epoch difference than GSC and AC used for STARNET, but PPM contains more
observations per star.
In general, we took the proper motions from STARNET except
in the following two cases:
(1) There was no entry in STARNET, maybe because the star was too
bright for the GSC or for any other reason. In our sample this was only
the case for SAO~76411~A, BD\,+17~724~B and BD\,+12~511. 
(2) The proper motions from STARNET and PPM deviated significantly from
each other. This happened for
the stars HD 283798 and BD\,+07 582(B), for which STARNET gives very high
proper motions. We assume that this is due to a wrong
identification of the star on the plates. This is a consequence of the very
large epoch difference, which on the one hand reduces errors in the proper
motions, but on the other hand makes the identification of high proper motion
stars more difficult and sometimes erroneous. In these cases, PPM is
the more reliable source, as it contains more observations per star.
Stars with large proper motions are discussed in
Sect.~\ref{starnet-discussion} in more detail.

The positions in all tables are valid for equinox and epoch J2000.0.
Note that the mean errors in
$\mu_{\alpha}$ are multiplied by $\cos\delta$ (whereas the proper
motions themselves are not).

\subsection{Proper motions of PMS stars known prior to ROSAT}

In STARNET or in the PPM
we could identify 34 PMS stars known prior to the ROSAT mission. Their
proper motions are given in Table~\ref{liste}. Of these, 15 stars had no
proper motion measurement so far. Previous determinations of
proper motions were performed in the surveys of Jones \& Herbig (1979)
(which was the most extensive one, including 80 suspected or confirmed T Tauri
stars together with 241 stars possibly associated with Tau-Aur) and in
the spatially more restricted surveys by Hartmann et al.\ (1991) and
Gomez et al.\ (1992). These previous papers publish relative proper
motions of stars on pairs of photographic plates. This is not an obstacle
to local kinematics. However, it prevents us from a comparison of the
individual accuracy in these previous papers and in the present one, because
the number of stars in common is so small.

\subsection{Proper motions of PMS stars discovered by ROSAT}

\subsubsection{The central region of Tau-Aur}
\label{center}
ROSAT observations in a region including the centre of Tau-Aur in combination
with optical follow-up spectroscopy have revealed 4~new classical and
72~new weak-line T Tauri stars (Wichmann et al.\ 1996). Of these, 38~stars
(all WTTS) have proper motions either in STARNET or in the PPM which are given
in Table~\ref{wich}.
Additional pointed ROSAT observations led to the discovery of 8 confirmed
(Strom \& Strom 1994), 2 likely and 5 possible TTS (Carkner et al.\ 1996), but
none of them could be found in STARNET because they are too faint.

\subsubsection{Region south of Tau-Aur}
\label{south}
Recently
Neuh\"auser et al.\ (1995b, 1997) and Magazz\`u et al.\ (1997) have studied
a sample of 111~stars in a region located just south of the Taurus 
molecular clouds. The field boundaries were chosen by these authors in such
a way that the Orion field (in the lower left in Fig.~\ref{pos}) was
excluded. The strip in the lower right of Fig.~\ref{pos}, which is 
perpendicular to the galactic plane, was included in
order to search for a gradient in the space density of TTS as a function
of distance from the Taurus clouds.

Proper motions of 58~stars in the sample of these authors are available, 
55 from STARNET, supplemented by 3 from PPM. 
The data are given in Table~\ref{neu}. 
Of these, 18~stars are classified as very young stars with ages 
$\le 3.5\cdot10^7$\,yrs (with lithium excess above Pleiades level), 
12 as stars with ages around $\approx 10^8$\,yrs
(with lithium, but no excess) and 28~stars as even older with ages
$> 10^{8}$\,yrs
(without any detected lithium) by Neuh\"auser et al.\ (1997).
The youngest stars with spectral types F or G are
just reaching the main-sequence, whereas stars with spectral
type K are pre-main sequence stars.

\section{Kinematics}

\subsection{Velocity dispersion}
Positions and proper motions of the stars from 
Tables~$\ref{liste}\mbox{-}\ref{neu}$
are shown in Figs.~\ref{pos} and \ref{pm}. We exclude stars with proper
motions clearly off the Taurus mean motion from the discussion in this
section; they are discussed separately in Sect.~\ref{starnet-discussion}.
Clustering of the remaining stars around the overall
mean values $(\mu_{\alpha}\cos\delta,\,\mu_{\delta})=
(4.0,\,-18.7)$~mas/y is visible in the proper motion
diagram, and  we immediately note that there is a relatively large scatter 
around the mean motion. Also, there is a slight
difference in motion between the stars in the central region
(Tables~\ref{liste} and \ref{wich}) with a mean proper motion of 
(2.4,\,$-21.1$)~mas/y and 
those from the southern region (upper part of Table~\ref{neu}) with a mean of 
(10.1,\,$-9.8$)~mas/y. The difference in $\mu_{\alpha}$\,($\mu_{\delta}$) 
is significant with confidence level larger than 95$\%$\,(99.9$\%$) in a 
$t$-test for distributions with different dispersions.

We first discuss the velocity dispersion inferred from
the scatter in proper motions. The distance of the Tau-Aur clouds
is determined to be 140\,pc (Elias 1978, Kenyon et al.\ 1994).
Recently Preibisch \& Smith (1997) have determined a best fit distance
of $152\pm10$\,pc on the basis of rotational properties of 25~WTTS, in
good agreement with previous determinations of the distance of the whole cloud.
At a distance of 140\,pc the scatter in proper motion corresponds to a velocity
dispersion in one coordinate of 6.5\,km\,s$^{-1}$ for the whole complex.
Splitting the sample into stars in the central and the southern
region, we find a velocity dispersion in the central part of Tau-Aur
of 5.4\,km\,s$^{-1}$, while the stars in the southern region
exhibit a velocity dispersion of 7.6\,km\,s$^{-1}$.

The mean error of the STARNET proper motions of 5\,mas/y (R\"oser 1996)
corresponds to 3.3\,km\,s$^{-1}$ at a distance of 140\,pc. Subtracting
this from the observed scatter
we get an intrinsic scatter of 4.3\,km\,s$^{-1}$ for the stars in the
central part and 6.8\,km\,s$^{-1}$ for the stars in the southern region.
These values appear very large compared to previous investigations.
Jones \& Herbig (1979) derive an overall intrinsic velocity dispersion
of 3.2\,km\,s$^{-1}$ and 2.2\,km\,s$^{-1}$ in their $x$- and $y$-direction 
($x$ essentially parallel to right ascension, $y$ parallel to declination),
respectively, but the region investigated by Jones \& Herbig (1979) is 
smaller than 
our 'central region' which is roughly the same as in Wichmann et al.\ (1996).

There is a significant difference in the determination of proper motions in
Jones \& Herbig (1979) and proper motions from STARNET. 
The region in Jones \& Herbig (1979)
is separated into subregions each corresponding to plate pairs.
Proper motions are determined differentially from these plate pairs. This
minimizes the effect of projection of the space motions over large areas
of the sky, which is inherent in our proper motions
because they are absolute proper motions (on the system of FK5).

\begin{figure}[t]
\epsfxsize=8.5cm
\leavevmode
\epsffile{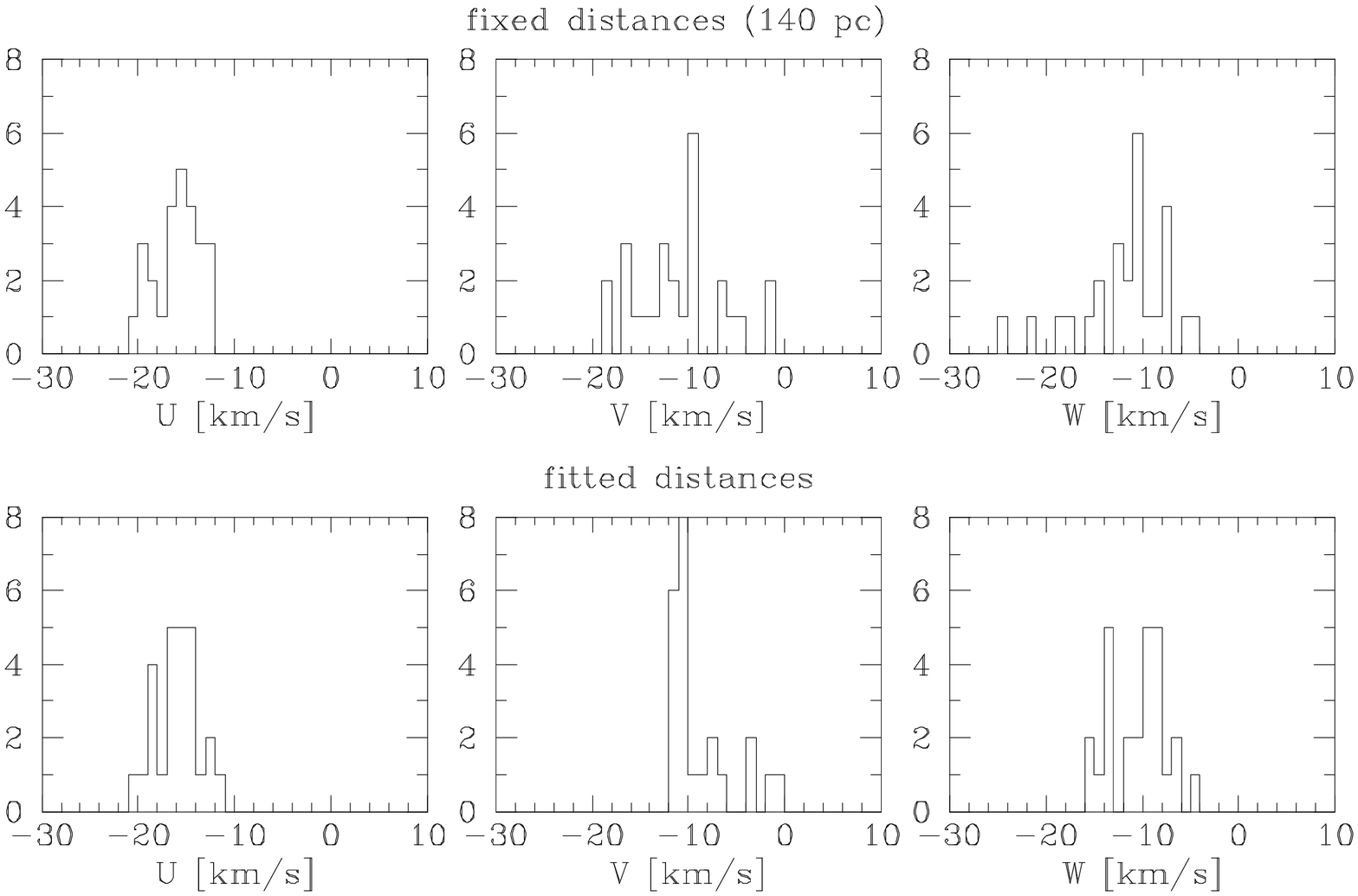}
\caption[]{\label{uvw} Space velocities for 26 TTS of Table~\ref{liste}
  with radial velocities available. In the upper panel space velocities are 
  calculated assuming fixed distances of 140\,pc for all stars,
  in the lower panel distances are adjusted so that the resulting
  space velocity for every star is as close as possible to the mean space
  velocity of the upper panel.
  $U$ points in the direction of the galactic centre, $V$ in the direction of
  galactic rotation and $W$ towards the north galactic pole.
  The highest peak in the $V$ velocities based on fitted distances 
  is due to 11 stars.}
\end{figure}

The mean proper motion of $\mu_x = 6.4$\,mas/y and $\mu_y =
-22.0$\,mas/y given by Jones \& Herbig (1979) is comparable to our mean
motion, indicating that the difference of the two astrometric systems 
is small. A thorough conversion of the proper motions between the
Jones-Herbig system of relative proper motions and the FK5 system of 
STARNET proper motions is not possible because only few stars are in common.

Jones \& Herbig (1979) subdivided the complex into smaller subregions, as they
had a fainter limiting magnitude and a larger number of stars.
Within these subgroups they determine the intrinsic velocity dispersion
in the following way. From the measured scatter of the proper motions they
subtract the scatter expected from the accuracy of their
measurements. However, these two 
quantities are almost equal. So, they derive an upper limit of 
1-2\,km\,s$^{-1}$ in most of their subgroups. The size of our sample
does not allow for a further subdivision. We can only study the velocity
dispersion of the complex as a whole. A velocity dispersion of
5.4\,km\,s$^{-1}$ would disintegrate the Tau-Aur complex in a time of the 
order $10^{7}$\,years, but typically smaller than the ages of
the PMS stars. We suppose that a large part of the measured velocity 
dispersion can be attributed to the ad hoc
assumption that all stars are situated at a distance of 140\,pc, and 
we discuss this in the next section.

\begin{figure}[t]
\begin{center}
\epsfxsize=6cm
\leavevmode
\epsffile{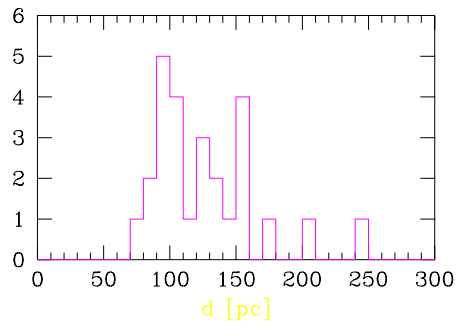}
\end{center}
\caption[]{\label{dist} Distance histogram for the stars shown in
  Fig.~\ref{uvw}. Distances are calculated requiring that the resulting
  space velocities are close to the mean. The mean of the distances is 
  127\,pc with a dispersion of 39\,pc.}
\end{figure}

\subsection{Space velocities}
As the stars of our sample populate a large region on the sky the
influence on
proper motions caused by projection effects has to be taken into account.
It is necessary to consider the total space velocities for a discussion of the
velocity dispersion. 
In the literature (see Table~1 in Neuh\"auser et al.\ 1995a for references) 
we found radial velocity measurements for 28~stars in the central area
of Table~\ref{liste}.

Space velocities of 26 of these stars (omitting the 2~stars with zero
proper motion components) calculated under the assumption of a fixed 
distance of 140\,pc for all stars are shown in Fig.~\ref{uvw} (upper panel).
The corresponding velocity dispersions are
$\sigma_{U} = 2.4\,$km\,s$^{-1}$,
$\sigma_{V} = 4.8\,$km\,s$^{-1}$ and
$\sigma_{W} = 4.7\,$km\,s$^{-1}$.
The low dispersion in $U$ is
caused mainly by the distance-independent radial velocities. The large
dispersion in $V$ and $W$ cannot be explained by the mean errors of STARNET
proper motions, and we suggest that it is (at least partly) due to
the lack of knowledge of the true distances.

We tested this hypothesis by varying the distance of each
star in order to minimize the difference between the corresponding space
velocity and the mean space velocity of the complex at a fixed distance
of 140\,pc. By this, the dispersion in the velocity components is significantly
reduced to 2.1\,km\,s$^{-1}$, 3.3\,km\,s$^{-1}$ and 2.8\,km\,s$^{-1}$
(Fig.~\ref{uvw}; lower panel).
The resulting velocity dispersion is now almost equal in all three
components, an indication for the correctness of our hypothesis.
A velocity dispersion of about
3\,km\,s$^{-1}$ in one component is very close to the formal error of
STARNET proper motions. This sets an upper limit to the intrinsic velocity
dispersion. This upper limit must be small compared to 3\,km\,s$^{-1}$
in order to have no influence on the measured velocity dispersion.
This result is consistent with the result of Jones \& Herbig (1979) for the 
smaller subgroups within their sample.

The distances calculated in the manner described above are shown in 
Fig.~\ref{dist}. The mean of these distances is 127\,pc, close to 140\,pc.
Furthermore our sample is biased towards brighter stars, i.~e.\ nearer
and/or earlier type stars, so a slightly lower value than the 
distance to the whole cloud complex is expected for these stars.
The dispersion of the distances is 39\,pc which is comparable to the
extent of the association in the tangential plane of at least 20\degr\ or
49\,pc at a distance of 140\,pc.

It is impossible to solve for the mean motion of the cluster
and the distances of the stars simultaneously.
Mi\-nimizing the velocity dispersion always favours lower
distances and a lower mean cluster motion, so all these values would tend
to zero. Minimizing only
the relative dispersion, normalized to the absolute value of the space
velocity, yielded a velocity dispersion which was much lower than
expected from the errors of the STARNET proper motions and in turn an
unbelievably high dispersion in the resulting distances.

These kinematically determined {\it individual} distances of the stars are 
not to be taken too literally; for this the method is too coarse.
However, the method yields a general tendency for the distribution of the
radial distances of the TTS in Taurus-Auriga.

\begin{figure}[t]
\epsfxsize=8.5cm
\leavevmode
\epsffile{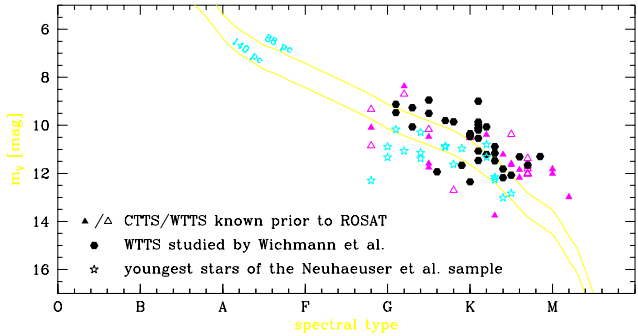}
\caption[]{\label{hr} Hertzsprung Russell diagram for the stars of 
  Tables~\ref{liste} and \ref{wich} and the youngest stars of Table~\ref{neu}
  without the high proper motion stars discussed separately in section~\ref
  {starnet-discussion}. Apparent $V$-magnitudes are taken from GSC 
  (typical errors 0.3\,mag)
  without corrections for extinction. Spectral types are taken from 
  SIMBAD for stars of Table~\ref{liste}, from Wichmann et al.\ (1996) for
  stars of Table~\ref{wich} and from Magazz\`u et al.\ (1997) for the stars
  of Table~\ref{neu}. The ZAMS is calculated for a distances of 140\,pc
  and 88\,pc, respectively. Some pre-ROSAT CTTS and WTTS appear below
  the ZAMS because we did not correct for absorption.}
\end{figure}

\subsection{Relation of the southern stars to Taurus-Auriga}

We discuss different scenarios for the origin of
the youngest stars in the southern region and their possible relation
to the Taurus molecular clouds.

(a) First we assume that the young stars in the southern region belong to the
Taurus-Auriga complex and share the same mean space motion. Because of their
different proper motions the southern stars cannot fulfill the requirement
of the same space motion if they are at a distance of 140\,pc. 
Varying their distances as described above (with a solution in the distance
interval between 50\,pc and 300\,pc for only 11 out of
16~stars), we find that they would be 
located at lower distances than the stars in the central area with a mean 
distance of 88\,pc.
This however leads to a conflict in
the HR diagram (Fig.~\ref{hr}): Nearly all the stars in the south would lie
below the main sequence for 88\,pc which is in contradiction to their
zero-age main sequence or even pre-main sequence nature. 
The velocity dispersions calculated with these distances are
$\sigma_{U} = 4.7\,$km\,s$^{-1}$,
$\sigma_{V} = 2.6\,$km\,s$^{-1}$ and
$\sigma_{W} = 3.7\,$km\,s$^{-1}$. The dispersions in $V$ and $W$ are close
to the values derived for the stars in the central region, but the
dispersion in $U$ (which is more or less independent of the distances) 
is higher than expected for members of a common star
forming region.

(b) If we assume that the stars are located at comparable distances to the
Taurus-Auriga association of about 140\,pc, this would make their location
in the HR diagram comparable to the Taurus member stars.
Kinematically they would not be related to the Taurus clouds, but the 
two complexes would rather approach each other with a relative velocity of
$\approx 9$\,km\,s$^{-1}$ and were adjacent to each other now only by chance.
It might be possible that the Taurus clouds originated in a high-velocity
cloud impact, so that the Taurus clouds oscillate around the galactic plane.
During the last passage through the plane, the stars were separated 
(combing-out) from the Taurus clouds, and now move ahead and already begin
to fall back to the plane. Alternatively, all the young stars south of
Taurus may just be Gould's belt members with typical ages of 3-5$\cdot
10^{7}$\,yrs; see Neuh\"auser et al.\ (1997) for a discussion.

(c) If the stars were more distant than 140\,pc the HR diagram would 
constrain them to be really very young pre-main sequence stars.
At the same time they would show a velcocity dispersion
higher than expected for a group of very young stars, and likewise 
very high X-ray luminosities.
In the tangential plane the two complexes would pass
more or less closely depending on the distance difference.

Neither of the above scenarios is completely convincing. Maybe the
stars in the southern area do not have a common origin and are not 
located at approximately similar distances.
Brice\~no et al.\ (1997) 
suggest that the population discoverd by ROSAT south of the Taurus clouds
is not made up of pre-main sequence but rather main-sequence stars
for which we do not expect that they share a common kinematic behaviour.
On the other hand only a small fraction of the youngest stars in 
Table~\ref{neu} are really PMS stars; about half of the PMS stars in
Neuh\"auser et al.\ (1997) are too faint for STARNET.
None of the younger stars appear to be ejected from the northern Tau-Aur
region. It should be kept in mind, however, that dynamically ejection
mechanisms favour low-mass escapers (Sterzik \& Durisen 1995). 
These stars are absent in this magnitude limited subsample.

Our kinematical findings indicate most probable that the PMS stars in the 
southern extension move towards the central Tau-Aur region. This implies
a larger separation of the two complexes in the past. From the kinematical
point of view a common star formation process seems therefore excluded.
The larger ages of the southern stars support our conclusion that
star formation in the two complexes must have been triggered by different
events.

\begin{table}[t]
\begin{center}
\caption[]{\label{test} Stars with conspicuous proper motions in our 
investigation. The last column indicates the result of the proper motion check
via POSS~I: a \raisebox{0.4ex}[0.4ex]{\tiny $\surd$}~-sign indicates that 
the proper motion could be 
confirmed, a ?-sign indicates a doubtful proper motion.}
\begin{tabular}{l@{\,}c@{$\;\,$}rrrrc@{\,\,}l}
\hline\noalign{\smallskip}
 & & \multicolumn{2}{c}{GSC No.\,/} & 
\multicolumn{1}{c}{$\mu_{\alpha}$} & \multicolumn{1}{c}{$\mu_{\delta}$} & \\ 
\raisebox{1.5ex}[-1.5ex]{object} & \raisebox{1.5ex}[-1.5ex]{Tab.} &
\multicolumn{2}{c}{PPM No.} & 
\multicolumn{2}{c}{[mas/y]} & \\
\noalign{\smallskip}\hline\hline\noalign{\smallskip}
\multicolumn{8}{l}{Pleiades candidates} \\
\noalign{\smallskip}\hline\noalign{\smallskip}
NTTS 032641+2420   & 1 & 1802 & 1190 &   31 & -43 & \raisebox{0.4ex}[-0.4ex]{\tiny $\surd$} & \\
NTTS 034903+2431   & 1 & 1804 &  123 &   27 & -51 & \raisebox{0.4ex}[-0.4ex]{\tiny $\surd$} & $^{(1)}$ \\
SAO 76411 A$\;^{(\ast)}$& 1 &  \multicolumn{2}{c}{93187} & 43 & -58 & \raisebox{0.4ex}[-0.4ex]{\tiny $\surd$} & $^{(2)}$ \\
RXJ 0407.9+1750    & 2 & 1254 &  785 &   18 & -49 & \raisebox{0.4ex}[-0.4ex]{\tiny $\surd$} & \\
RXJ 0409.2+2901    & 2 & 1826 &  877 &   39 & -37 & \raisebox{0.4ex}[-0.4ex]{\tiny $\surd$} & \\
HD 285579          & 2 & 1251 &  201 &   13 & -50 & \raisebox{0.4ex}[-0.4ex]{\tiny $\surd$} & \\
RXJ 0437.5+1851    & 2 & 1274 & 1515 &   16 & -52 & \raisebox{0.4ex}[-0.4ex]{\tiny $\surd$} & \\
RXJ 0439.4+3332A   & 2 & 2378 & 1232 &   30 & -40 & \raisebox{0.4ex}[-0.4ex]{\tiny $\surd$} & \\
RXJ 0448.0+0738    & 3 &  683 &  661 &   25 & -59 & ? & \\
HD 287017          & 3 &  687 &  419 &   52 & -73 & \raisebox{0.4ex}[-0.4ex]{\tiny $\surd$} & $^{(2)}$ \\
\noalign{\smallskip}\hline\noalign{\smallskip}
\multicolumn{8}{l}{other high proper motion stars} \\
\noalign{\smallskip}\hline\noalign{\smallskip}
RXJ 0210.4-1308 SW & 3 & 5283 & 1690 &   55 & -24 & \raisebox{0.4ex}[-0.4ex]{\tiny $\surd$} & \\
RXJ 0212.3-1330    & 3 & 5283 &  876 &  163 & -81 & \raisebox{0.4ex}[-0.4ex]{\tiny $\surd$} & $^{(2)}$ \\ 
HD 15526           & 3 & 5284 &  686 &   48 & -13 & \raisebox{0.4ex}[-0.4ex]{\tiny $\surd$} & \\
RXJ 0239.1-1028    & 3 & 5288 & 1027 &   -8 & -95 & \raisebox{0.4ex}[-0.4ex]{\tiny $\surd$} & \\
RXJ 0317.9+0231    & 3 &   59 &   24 &  -15 & -61 & ? & \\
RXJ 0330.7+0306 N  & 3 &   67 &  206 &   33 & -85 & ? & \\
RXJ 0336.0+0846    & 3 &  657 &  726 &    1 &  26 & \raisebox{0.4ex}[-0.4ex]{\tiny $\surd$} & \\
RXJ 0403.4+1725    & 2 & 1254 &  309 &  -66 & -15 & ? & \\
BD\,+00 760        & 3 &   75 & 1529 &   57 &  -8 & \raisebox{0.4ex}[-0.4ex]{\tiny $\surd$} & $^{(2)}$ \\
LkCa 14            & 1 & 1834 &  177 & -161 &  97 & ? & \\
RXJ 0523.9+1101    & 3 &  704 & 2073 &    2 &-115 & ? & \\
\noalign{\smallskip}\hline\\[-0.3ex]
\end{tabular}
\begin{footnotesize}
\begin{tabular}{c@{$\;$}p{8cm}}
$^{(\ast)}$ & classified as non-member by van Leeuwen et al.\ (1986) \\[0.3ex]
$^{(1)}$    & checked by comparision with Schilbach et al.\ (1995) \\
$^{(2)}$    & proper motion from PPM / checked in PPM \\
\end{tabular}
\end{footnotesize}
\end{center}
\end{table}

\section{High proper motion stars in STARNET}
\label{starnet-discussion}
\subsection{Verification of high proper motions}
Prior to the physical interpretation of the large proper motions of a number
of stars, these proper motions have to be carefully investigated. As
mentioned, STARNET proper motions are derived from AC and GSC.
For a certain number of stars in STARNET misidentifications of their positions
on the AC and/or GSC plates can occur because of the large 
($\approx 80$\,years) epoch difference.
We decided to cross-check large proper motions either by comparison
with the proper motion given in another catalogue, if available, or by
testing the star's position on the plates of a third epoch.
This is by no means an easy task, because accurate astrometric data in this
magnitude range are rare.

Table~\ref{test} summarizes the stars with large proper motions from 
Tables~\ref{liste}, \ref{wich} and \ref{neu}.
The sample can be split into stars with proper
motions close to the mean proper motion of the Pleiades cluster
($\mu_{\alpha}\approx 16$\,mas/y, $\mu_{\delta}\approx -44$\,mas/y) and those
with proper motions randomly far off the mean motion of Tau-Aur.
The proper motions of 15 stars in Table~\ref{test} could be confirmed
by comparison with their positions on the digitized POSS~I plates or by
independent proper motion measurements. This is  indicated
by a \raisebox{0.4ex}[-0.4ex]{\tiny $\surd$}~-sign. 
The digitized POSS~I is well suited to check the erroneously large
proper motions because in this case
there should be a large, easily detectable offset from the expected position.
This turned out to be the case for the 6 remaining stars; their proper motions
seem to be erroneous in STARNET

\subsection{Pleiades membership}
In the proper motion plot (Fig.~\ref{pm}) we find a secondary crowding
of stars with proper motions similar to that of the Pleiades.
Indeed the star NTTS~034903+2431 is found to match both photometric and
proper motion membership criteria of the Pleiades cluster by Schilbach et
al.\ (1995, their star No.~36000) and classified as highly probable
Pleiades member. SAO~76411~A, on the other hand, corresponds to the star 
Pels 178, which was included in the
photometric investigation of the Pleiades cluster by van~Leeuwen et al.\ (1986)
and classified as non-member.

\begin{figure}[t]
\begin{center}
\epsfxsize=7.5cm
\leavevmode
\epsffile{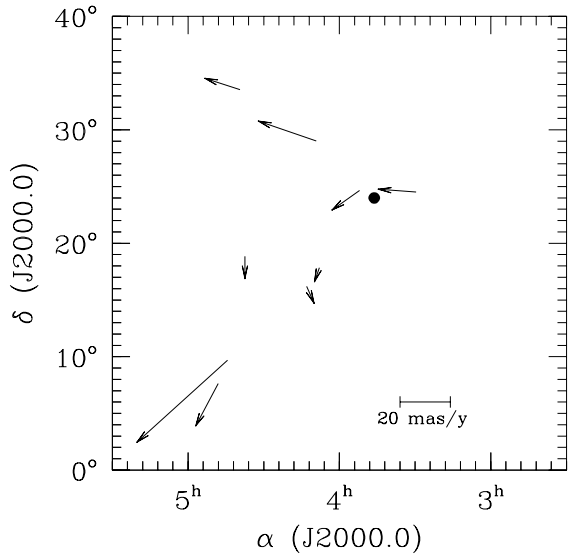}
\end{center}
\caption[]{\label{ple} Position of the Pleiades member stars and direction
  of their proper motions relative to the cluster mean 
  ($\mu_{\alpha}=16$\,mas/y, $\mu_{\delta}=-44$\,mas/y).
  The filled circle
  indicates the centre of the Pleiades cluster at about $\alpha=3\,\fh8,\;\: 
  \delta=24\,\fdg0$\,. The
  scale for the length of the arrows is indicated in the lower right corner
  of the plot.}
\end{figure}

Figure~\ref{ple} shows the distribution on the sky of the stars from the
upper part of Table~\ref{test} with the exception of SAO~76411~A.
The direction to the centre of the Pleiades cluster is indicated by a
filled circle. The arrows show the proper motions of the stars from
Table~\ref{test} relative to the mean motion of the Pleiades in PPM and
STARNET. Two stars close to the centre of the Pleiades have small motions
relative to the Pleiades cluster; one of them is the star classified
as highly probable proper motion member by Schilbach et al.\ (1995).
Three stars about 10\degr\ away from the Pleiades centre show no relative
motion and should be checked further photometrically for membership.
Finally, the stars RXJ~0409.2+2901, RXJ~0439.4+3332A, RXJ~0448.0+0738 and
HD~287017
could have been ejected from the Pleiades a few million years ago.
This is consistent with the findings of Kroupa (1995), who expects from 
the result of numerical N-body simulations of star clusters that
during the lifetime of a cluster a certain fraction of stars can be ejected
with velocites up to 100\,km/s due to close encounters between binary systems. 
The ejection rate is estimated to be higher in the earlier phases of cluster 
evolution.

\section{Summary and conclusion}
We have presented new proper motions taken from the STARNET catalogue of
62~stars in the central region of Taurus-Auriga whose kinematics is 
consistent with membership to the association. 
These stars show a velocity dispersion small compared to 3\,km\,s$^{-1}$,
a limit which is given by the accuracy of the STARNET proper motions.
This is consistent with the derivation of an
upper limit for the velocity dispersion in subgroups of the Taurus-Auriga
complex of about 1-2\,km\,s$^{-1}$ by Jones \& Herbig (1979).

Among the high proper motion stars in our sample we found 8~new Pleiades 
candidates or runaways
(and re-identified 1~known before). They are located up to $\approx 20\,\degr$
away from the centre of the Pleiades cluster, and their directions of motion
are consistent with them being ejected from the cluster a few million
years ago.

Finally, all attempts to kinematically relate the recently detected PMS
stars south of the Taurus-Auriga cloud to the central part of this
association have failed. 
We do not find any young stars in our sample that appear to be ejected
{}from the Taurus-Auriga association.
We conclude from this that the star formation scenario
- at least for the G and early K type stars -
in this southern area is different from the one in the central part.

\begin{acknowledgements}
This research has made use of the SIMBAD database, operated at CDS,
Strasbourg, France, and of the APS Catalog of the POSS~I
which is supported by the National Science Foundation, the National
Aeronautics and Space Administration, and the University of Minnesota.
The APS databases can be accessed at http://isis.spa.umn.edu\,.\\
S.F., R.N.\ and M.F.S.\ acknowledge grants from the Deutsche 
Forschungsgemeinschaft (DFG Schwerpunktprogramm `Physics of star formation').
\end{acknowledgements}

\end{document}